# Chromium-doped silica optical fibres : influence of the core composition on the Cr oxidation states and crystal field

Authors: V. Felice, B. Dussardier, J. K. Jones, G. Monnom, D. B. Ostrowsky

Affiliation: Laboratoire de Physique de la Matière Condensée (UMR CNRS-UNSA n° 6622), Université de Nice Sophia Antipolis, Parc Valrose, 06108 Nice cedex 2, France

Corresponding author:

B. Dussardier

LPMC-Université de Nice Sophia-Antipolis, Parc Valrose, 06108 Nice Cedex 2, France

Tel : (33) 4 92 07 67 48, fax : (33) 4 92 07 67 54

Email address : Bernard.Dussardier@unice.fr

## ABSTRACT

We have made chromium-doped, silica-based preforms and optical fibres by Modified Chemical Vapour Deposition and have studied the influence of the chemical composition of the doped region on the Cr-oxidation states and the spectroscopic properties of the samples. Chromium, introduced initially as $Cr^{3+}$, is partially or totally oxidized during the fabrication process, and is stabilized in the core of the preforms and fibres as $Cr^{3+}$ in octahedral coordination, and/or as $Cr^{4+}$ in distorded tetrahedral coordination ($C_s$). Small concentrations (~1-2 mol%) of codopants in silica, such as germanium or aluminium, suffice to promote a particular oxidation state. Particularly, 1 mol % of aluminium stabilizes all present chromium into $Cr^{4+}$. The ligand field parameters, *Dq, B* and *Dq/B*, for $Cr^{3+}$ and $Cr^{4+}$ are derived. It is shown that the chromium ions in our samples are in low to intermediate ligand field and that *Dq* and *Dq/B* decrease when the aluminium content increases. The absorption cross sections of $Cr^{3+}$ are similar to that reported in glasses and crystals, whereas $Cr^{4+}$ has lower values than reference laser materials.

## 1. Introduction

Optical fibre materials with very broad-band gain are of great interest for developping widely tunable miniature lasers and amplifiers for telecommunications with a wider gain band than commercial erbium-doped fibre amplifiers and other reported rare earth (RE)-doped fibre devices. Tunability in RE-doped fibre lasers is already well established [1], but is limited to few tens of nanometers due to the intrinsic narrowness of the RE optical transitions. Besides, bulk solid-state lasers materials, doped with chromium, such as $Cr^{4+}$:YAG, have demonstrated very good results as broad-band gain media [2], and also as saturable absorbers [3,4]. Therefore, it is of great interest to develop transition metal (TM) ion-doped optical fibres in order to implement their optical properties (as amplifying media or saturable absorbers) into optical fibre-based devices and systems. We have reported for the first time on $Cr^{4+}$-doped silica optical fibres, which in particular exhibited a very broad (>500 nm) luminescence in the near infrared (NIR), centrered at 1250 nm [5]. The choice of vitreous silica for the fibre material is of critical importance, as all operating systems are based on silica fibre technology. Little literature exists on chromium- and other TM-doped vitreous bulk silica, although this issue has already been addressed in the 70's [6] with the aim of lowering attenuation in silica optical fibres. Some reports on chromium-doped glasses have already shown evidence of absorption and NIR fluorescence due to $Cr^{4+}$ in these materials [7,8]. However the techniques used to prepare them and their compositions greatly differ from those of silica optical fibres. Therefore, some basic studies on the optical properties of TM ions in silica-based optical fibres are needed. In particular, the final oxidation state(s) present in the fibre core depends strongly on the preparation technique and some fabrication parameters. Also, the optical properties of one particular oxidation state of a TM ion varies from one host composition and structure to another, due to crystal-field (or so-called ligand field) variations [9]. These various effects render the interpretation of absorption and emission spectroscopy difficult. The aim of this paper is to address some of these issues in the case of chromium, as this element has the best potential for applications.

In this paper, we report on $Cr^{3+}$ and $Cr^{4+}$ ions in silica preforms and fibres, prepared by the Modified Chemical Vapour Deposition (MCVD) and doped by the so-called ‘solution doping technique’. We show that by changing the core composition of the preform, particularly its content in aluminium and germanium, we can preferentially stabilize $Cr^{3+}$ and/or $Cr^{4+}$ in the preform core. In particular, we show that in an aluminosilicate fibre core (containing only 1% of aluminium), all the present chromium is stabilized into the 4+ valence state. This has very interesting application prospects, either as amplifying medium or as saturable absorber in ‘all-fibre’ devices. Also, by slightly varying the ion environment composition, we can modify the ligand field, as seen from changes in the absorption spectra and parameters such as the ligand field *Dq* and the normalized lignand field *Dq/B*. The absorption cross-sections of $Cr^{3+}$ and $Cr^{4+}$ are calculated and compared to literature.

## 2 Experimental

### 2.1 Preforms and fibre fabrication

The silica preforms studied here were fabricated by MCVD [10] and were doped with chromium by the classical ‘solution doping technique’, originally developped for RE-doping [11]. During the MCVD process stage, the silica layers which will eventually constitute the core of the fibre were deposited at a lower temperature than the preceding cladding layers, so that they were not fully sintered and left porous. Then the substrate tube was filled with an alcoholic solution of a $Cr^{3+}$-salt ($CrCl_3$:6 $H_2O$) and allowed to impregnate the porous layers. After a few hours, the tube was emptied, the layers were dried and fully sintered under nitrogen, and subsequently the substrate tube was collapsed into a 10 mm-diameter preform in the usual way. In the preform, the doped region (the core) diameter is typically 0.5-1.0 mm.

Germanium and aluminium were used to rise the fibre core refractive index. Germanium was incorporated during the deposition stage, by MCVD, while aluminium was introduced during the soaking stage (from an $Al^{3+}$-salt, $AlCl_3$:5 $H_2O$, dissolved in the soaking solution). All the samples contain a small concentration of phosphorus (~0.5 mol%), acting as a glass softener. Three different types of core compositions were prepared, containing only germanium, germanium and aluminium, and only aluminium as refractive index riser. They are referred to as Cr(Ge), Cr(Ge-Al) and Cr(Al), respectively. For the purpose of this study, the preforms were doped using soaking solutions with a relatively high chromium strength. A Cr(Al)-type preform was also prepared with a low chromium concentration solution, and pulled into a fibre having a 10 µm-diameter core.

### 2.2 Sample core composition and chromium oxidation state characterization

The chemical composition of each preform core was determined by Plasma Emission Spectroscopy (PES). The core was extracted from the preform by sawing, and then ground before analysis. The nominal sensitivity of this technique (1 ppm) was degraded to ±15% because of the effective core volume estimation. The PES results (Table 1) show that the silica-based core composition is only slightly modified by so-called ‘co-dopants’ such as Ge, Al and P, and that the ratio $[Cr_{total}]$/[co-dopants] is always less than unity. All the preforms having chromium contents higher than several hundreds molar parts per million (mol ppm) show evidence of phase separation: under optical microscope, radial Cr-rich (highly green or blue coloured) regions are seen amidst less coloured transparent glass. The low Cr-concentrated preform (designed for making the fibre) has an estimated concentration of 40 ppm, and does not show any phase separation. This shows that doping silica fibres with high-strength Cr-solutions causes separation of the core material into Cr-rich phases and silica-rich phases, as was observed in highly rare-earth-doped silica fibres [12]. Homogeneously doping these fibres with higher chromium concentrations would be possible only if the silica network was modified with higher concentrations of codopants, like aluminium, phosphorus or boron for instance.

The oxidation states of chromium present in the samples were analyzed by Electron Paramagnetic Resonance (EPR). The measurements were performed on core samples of identical volume extracted from Cr(Ge), Cr(Ge-Al) and Cr(Al) prefoms (Figure 1). The Cr(Ge) and Cr(Ge-Al) EPR spectra present two well defined bands noted A and B and a band at lower field value, noted C, larger and less well defined than bands A and B. On the other hand, only bands B and C are visible on the spectrum of Cr(Al). On Table 2 are reported the magnetic field and intensities of the EPR bands for the three samples.

**2.3 Absorption measurements**

Absorption measurements on preforms were performed at RT on 2 mm-thick slices. The beam of a white light source was chopped and focused into the core of the sample. The transmitted light was collected by a 100 μm-core optical fibre and directed into a motorized monochromator (2 nm steps, resolution 40-80 $cm^{-1}$). Visible and NIR output light was detected by a Si or Ge detector, respectively, and a lock-in amplifier synchronized on the chopper frequency. Absorption spectra were corrected from the apparatus spectral response by performing a second measurement through the cladding of the preform. Several measurements were performed on each sample, giving a ±15 % error on the absorption coefficient. The results for Cr(Ge), Cr(Ge-Al) and Cr(Al) are shown in Figure 2. As expected from the phase-separated nature of the samples, the absorption spectra show strong background losses attributed to light scattering. However, the bands due to absorption by chromium ions are visible and can be extracted from the background losses (see below).

The absorption spectrum of the preform Cr(Ge) (Fig. 2(a)) presents two bands around 15000 $cm^{-1}$ and 21000 $cm^{-1}$. The absorption spectrum of the preform Cr(Ge-Al) (Fig. 2(b)) is characterized by a broad absorption band in the visible to NIR range, from 19000 downto 10000 $cm^{-1}$. A weak shoulder around 9000 $cm^{-1}$ on the low energy side of the broad band and a band near 21000 $cm^{-1}$ are seen. Above 22000 $cm^{-1}$, the spectrum resolution is degraded by strong losses and high noise, and is not shown here.

The Cr(Al) absorption spectrum was measured in the low-Cr doped fibre. The transmitted light was detected with the same setup as for the preform samples. The absorption spectrum was corrected from the apparatus spectral response by performing a second measurement with a shorter length of fibre. The absorption spectrum (Fig. 2(c)) main features are a very broad absorption band in the visible to NIR range, from 20000 $cm^{-1}$ downto 10000 $cm^{-1}$, and a shoulder around 9000 $cm^{-1}$ on the low energy side of the broad band. Note that the background losses due to scattering are negligible compared to those in the highly doped preforms. Absorption measurements in the highly doped Cr(Al) preform shown the same absorption shape, but superposed on strong scattering losses. Low temperature (77 K) measurements were also performed on this fibre. Although the spectrum was slightly shifted toward higher energies, from room to low temperature, there was no noticeable difference between both spectra [5]. We observed also this quasi-insensitivity of the absorption to temperature in fibres with different core compositions, in agreement with Ref.[8] in alumino-silicate glasses.

## 3 Interpretation and discussion

### 3.1 EPR spectra

On the EPR spectra of Cr(Ge) and Cr(Ge-Al), band A is assigned to $Cr^{3+}$, while band B and band C, which is at a field value approximately the half of the value of band B, are assigned to $Cr^{4+}$ ions [13]. Therefore, in these samples, both $Cr^{3+}$ and $Cr^{4+}$ ion oxidation states are present. No other chromium oxidation states is detectable. This suggests that during the doping and final MCVD process stages, the $Cr^{3+}$ ions introduced by the soaking solution were partially or totally oxidized into $Cr^{4+}$ or higher oxidation states. However, besides $Cr^{3+}$, only $Cr^{4+}$ was stabilized in our samples. In particular, $Cr^{6+}$ compounds may have been synthesized but were volatilized during the preform collapse stage. As the Cr(Al) EPR spectrum displays only bands B and C, and not band A, only $Cr^{4+}$ was stabilized in the preform. This result is of particular interest for applications operating in the NIR range.

As the EPR measurements were performed on samples having the same core volume, the intensity of each EPR band (relative to the base line of the spectra) is proportional to the quantity of $Cr^{3+}$ or $Cr^{4+}$ ions. With the values obtained from Cr(Ge) and Cr(Ge-Al) on bands A and B, we calculated the relative concentration of $Cr^{3+}$ and $Cr^{4+}$ ions present in theses samples. From the total chromium concentration, obtained by PES, the absolute $Cr^{3+}$ and $Cr^{4+}$ concentrations were also calculated. These results are also listed in Table 2. Within the uncertainty of the composition results, the $[Cr^{4+}]/[Cr_{total}]$ ratio increases lineraly with [Al] and reaches 100 % when germanium is totally absent, and [Al] = 1.0 mol % only. This behaviour suggests that the $Cr^{3+}$ ions are preferentially located in Ge-modified sites, and the $Cr^{4+}$ ions are preferentially stabilized into Al-modified sites in the glass.

### 3.2 Absorption spectroscopy

We have tentatively interpreted the absorption spectra in order to assign each observed absorption band to a particular oxidation state of chromium. This was performed in the goal of confirming the above results and conclusion from the EPR measurements, and also to estimate the respective absorption cross sections of $Cr^{3+}$ and $Cr^{4+}$ in the silica-based preforms and optical fibres.

Each absorption spectrum was corrected from the scattering background losses and decomposed into a sum of appropriate Gaussian curves. Indeed, the fluorescence and absorption respective behaviour versus temperature [5] can be interpreted from a simple configuration coordinate model, with vibrationally broadened optical transitions [9]. Therefore, a single transition bandshape is approximately a Gaussian, and the global absorption spectrum is a sum of Gaussian curves. The curves parameters were fitted to the experimental data points with a good adjustment. In particular, the peak energies ($E$) and widths ($\Delta E$) are found well within a ±100-$cm^{-1}$ error. The ±15 % error on the absorption coefficients ($\alpha$) was mainly introduced by measurements errors, and not by the fitting procedure. The parameters found for Cr(Ge) are listed in Table 3. The corresponding background-corrected absorption spectrum and the obtained Gaussian curves are shown in Figure 3. The two absorption bands are assigned to $Cr^{3+}$ in octahedral symmetry. Their characteristics are very similar to those found for $Cr^{3+}$ in other oxide glasses [6-8,14,15]. The spin forbidden $^4A_2 \rightarrow {}^2E$ transition is expected to be very weak, and was not observed. Also, no absorption band was found for $Cr^{4+}$, although this oxidation state is present in this sample. This will be discussed later. The same procedure was applied to the absorption spectra of the preform Cr(Ge-Al) and the fibre Cr(Al). The corresponding parameters are listed in Table 3 and the corresponding background corrected absorption and fitted curves are shown in Figures 4 and 5, respectively. The Cr(Al) spectrum was treated first, because all the bands could be assigned to $Cr^{4+}$. They are tentatively assigned to electronic spin-allowed transitions of $Cr^{4+}$ in distorded tetrahedral symmetry Cs [16,17]. The distortion of the perfect tetrahedral symetry partially lifts the degeneracy of some energy levels such as $^3T_1(^3F)$ and causes the presence of several peaks assigned to one transition. The Cr(Al) absorption spectrum is similar to those observed in aluminate [7] and alumino-silicate glasses [8], in which there is evidence of the removal of the orbital degeneracy of the $^3T_1(^3F)$ state. In the following, the $Cr^{4+}$ energy states are referred to by their irreductible representation in the non distorded tetrahedral ($T_d$) symmetry coordination, were the ground state is $^3A_2(^3F)$. As expected, the spin forbidden $^3A_2 \rightarrow {}^1E$ weak transition was not observed.

Then, the absorption spectrum of Cr(Ge-Al) was decomposed in the light of the results on Cr(Ge) and Cr(Al). When comparing the arrangement of the $Cr^{3+}$ and $Cr^{4+}$ bands found in Cr(Ge-Al) with those found in Cr(Ge) and Cr(Al), respectively, a good qualitative agreement is found, although some slight discrepancies can be seen. In particular, the width of the $^4A_2(^4F) \rightarrow {}^4T_1(^4F)$ transition of $Cr^{3+}$ in Cr(Ge-Al) is much lower than that in Cr(Ge). This is caused by the curve fitting range being limited to < 21500 $cm^{-1}$, due to the bad absorption spectrum resolution above this energy. Therefore the curve fitting could not resolve for the high energy side of the Gaussian curve found for this transition. Also, the arrangement and widths of the three Gaussian curves found for the $Cr^{4+}$:$^3A_2(^3F) \rightarrow {}^3T_1(^3F)$ transition in Cr(Ge-Al) differs from that in Cr(Al). In preform Cr(Ge-Al), the strong $Cr^{3+}$ 14440 $cm^{-1}$-band partially masks the $Cr^{4+}$ 12310 $cm^{-1}$-band and hampers its precise peak energy fitting. However, the mean energy of the $Cr^{4+}$:$^3A_2(^3F) \rightarrow {}^3T_1(^3F)$ transition (calculted as the first order momentum of the sum of the three Gaussian curves) is found at 13710 $cm^{-1}$, which compares well with that found in Cr(Al) at 13900 $cm^{-1}$. We observe also that the $Cr^{4+}$:$^3A_2(^3F) \rightarrow {}^3T_1(^3F)$ transition bands are larger (above 3000 $cm^{-1}$) in Cr(Al) than in Cr(Ge-Al) (around 2250 $cm^{-1}$). This may be an effect of the higher aluminium concentration in Cr(Al) than in Cr(Ge-Al), causing a larger silica network dislocation and therefore offering a larger variety of sites in the host. This causes inhomogeneous broadening of some transitions sensible to ligand field variations, such as $Cr^{4+}$:$^3A_2(^3F)\rightarrow{}^3T_1(^3F)$. This effect is well known in rare earth-doped, aluminium-codoped silica optical fibres [18,19].

In order to compare our optical spectroscopic results with other chromium-doped glasses reported in the literature (although none has a composition close to our silica-based samples), we have derived the ligand field strength *Dq*, the Racah parameter *B* and the ratio *Dq/B* for both $Cr^{3+}$ and $Cr^{4+}$ in the three samples, from the data in Table 3, and using the formalism of Tanabe and Sugano [9,20] The results are listed in Table 4. We calculated the energy of the split $Cr^{4+}$:$^3A_2(^3F)\rightarrow{}^3T_1(^3F)$ transition as the first order momentum of the sum of the three Gaussian curves. The uncertainties listed in Table 4 are maximum values; they were calculated on the basis of a maximum error of ± 100 $cm^{-1}$ for the peak energy, estimated from the monochromator resolution and the curve fitting procedure. When considering *Dq* and *Dq/B* values for $Cr^{3+}$ and $Cr^{4+}$ in Table 4, one sees that chromium ions are in low-to-intermediate ligand field sites, in agreement with reports on other oxide glasses, either silicates doped with $Cr^{3+}$ or alumino-silicates doped with $Cr^{4+}$ (from [8]). In our samples, $Cr^{3+}$ ions lie in the so-called 'crossing' region of the $^4T_2$ and $^2E$ states (at which Dq/B~2-2.2) like in silicate glasses [14,15], but *Dq* and *B* are found ~90 $cm^{-1}$ (6 %) and ~70 $cm^{-1}$ (10 %) (in average) lower than in these glasses, respectively. This would suggest that $Cr^{3+}$ ions are submitted to a lower ligand field in the silica fibres than in classically prepared glasses, and that the covalency of their bonding to the glassy network is increased [20]. Furthermore, $Cr^{3+}$ in pure bulk silica is submitted to rather different ligand field conditions than in our samples (*Dq* = 1600 $cm^{-1}$, *B* = 560 $cm^{-1}$ and *Dq/B* = 2.88, from [6]). This shows that glass network builders, like Ge, and modifiers, like Al and P, induce great spectroscopic properties changes, although these elements are present in low concentrations (~1-2 mol%).

In the case of $Cr^{4+}$, *Dq/B* is found equal or lower than the 'crossing region' value (*Dq/B* ~1,6), particularly in the Cr(Al) sample. To our knowledge, no data on *B* and *Dq/B* in glasses has been reported in the literature. However, it is interesting to compare our results with those reported in $Cr^{4+}$-doped laser crystals, like YAG and forsterite [17 and references therein]. In these materials, *Dq* ~ 1000 $cm^{-1}$, *B* ~ 645 $cm^{-1}$ and therefore *Dq/B* ~ 1.55. These values are quite similar to our results, although in our samples, *Dq* is found 10 % lower. In that situation, a broad luminescence is expected along the $^3A_2(^3F)\rightarrow{}^3T_1(^3F)$ transition, as in YAG and forsterite laser materials, and was observed in the Cr(Al) fibre [5]. Although the *Dq*, *B* and *Dq/B* values found in our samples series are inherently approximate (due to the width and overlap of the absorption bands), a trend in their variations can also be seen. As the aluminium content is increased, and the germanium decreased, *Dq* and *Dq/B* decrease, for both $Cr^{3+}$ and $Cr^{4+}$, while *B* remains within its uncertainty. However, this trend is not conclusive, and would need more investigations, particularly by further modifying the silica network around the active ions.

From the PES, EPR and absorption measurements, the $Cr^{3+}$ and $Cr^{4+}$ absorption cross sections were calculated for each of their assigned transition. The results are reported in Table 3. The estimated uncertainties due to the measurements are ± 45 % for Cr(Ge) and Cr(Ge-Al), and ± 30 % for Cr(Al). The values found for $Cr^{3+}$ in Cr(Ge) and Cr(Ge-Al), and for $Cr^{4+}$ in Cr(Ge-Al) and Cr(Al), respectively, are equal within their. The absorption cross section of the $Cr^{3+}$:$^4A_2(^4F)\rightarrow{}^4T_2(^4F)$ (~680 nm)-peak in our samples compares well with that estimated in pure silica (from [6], after correction of the absorption spectrum from the strong UV-absorption tail of $Cr^{6+}$) for the same transition at 625 nm (σ ~ $30.10^{-24}$ $m^2$), and in ruby (σ ~ $20.10^{-24}$ $m^2$) [21]. However, the absorption cross-section of $Cr^{4+}$ around 1100 nm in our samples are found lower than in $Cr^{4+}$-doped laser crystals, by one or two orders of magnitude [3]. The only comparison with $Cr^{4+}$-doped glasses that can be done is with aluminate [7] and alumino-silicate glasses [8]. Although theses reports did not give actual absorption cross-section values around 1100 nm for $Cr^{4+}$, they can be estimated to ~$40.10^{-24}$ $m^2$ and ~$5.10^{-24}$ $m^2$, respectively. Our values (~$3.10^{-24}$ $m^2$) are closer to the alumino-silicate than the pure aluminate one, as expected in a aluminium-modified silica network. Also, $Cr^{4+}$ has cross-sections an order of magnitude lower than $Cr^{3+}$ in the silica-based preforms. This explains why the contribution of $Cr^{4+}$ is not detected in the Cr(Ge) absorption spectrum, although this valency constitutes 36 % of the total chromium content.

## 5 Conclusion

In three different types of chromium-doped silica-based preforms and optical fibres, we have assigned the absorption bands to spin allowed electronic transitions of two oxidation states of chromium ions. We have also estimated the respective absorption cross-sections of these transitions. By slightly modifying the concentration in germanium and/or aluminium (of less than 2 mol%) in the core of the samples, their optical properties are greatly modified. In particular, we show that:

i) Only the $Cr^{3+}$ and $Cr^{4+}$ oxidation states are stabilized in the samples, in octahedral site symmetry and distorded tetrahedral site symmetry, respectively.

ii) Germanium favours predominantly the $Cr^{3+}$ oxidation state, whereas aluminium offers sites for $Cr^{4+}$. In particular, the sample co-doped with only aluminium (1 mol%) contains only $Cr^{4+}$. This is of great interest for applications in fibre devices working in the NIR range, such as passively Q-switched fibre lasers using $Cr^{4+}$ as saturable absorber.

iii) Increasing the aluminium concentration lowers the normalized ligand field *Dq/B* applied on the chromium ions. If this trend is confirmed by further investigations, this would be a mean of engineering the optical properties of silica-based chromium-doped fibres.

iv) The absorption cross sections of $Cr^{3+}$ are unchanged when varying the core composition in Ge and Al. The same is observed for $Cr^{4+}$. $Cr^{3+}$ cross sections (~$35.10^{-24}$ $m^2$) are consistent with reported values in other materials, while the $Cr^{4+}$ cross sections (~$4.10^{-24}$ $m^2$) are lower than in reference laser crystals, but consistent with those reported in glasses.

v) Phase separation occurs for chromium concentrations higher than several hundreds mol ppm in our series of samples. Higher dissolution would be achieved by further modifying the silica network. However, because we are dealing with optical fibres, low Cr-concentrations, such as 40 mol ppm or less, can be used in long interaction lengths.

**Acknowledgements**

The authors acknowledge D. Simmons, J. Théry and D. Vivien, from the Laboratoire de Chimie Appliquée de l'Etat Solide (ENSCP, Paris) for the EPR measurements and interpretation, and also for helpful discussions. J.K. Jones was supported by the European Communities Council (TMR program, no. ERBFMBICT960647).

**Table and figures captions**

**Tables :**

1 Core composition of the preforms Cr(Ge), Cr(Ge-Al) and Cr(Al). Uncertainty ±15 %.

2 Position and intensity of the EPR bands in the preforms Cr(Ge), Cr(Ge-Al) and Cr(Al), and their respective contents in $Cr^{3+}$ and $Cr^{4+}$.

3 Energy ($E$), full-width at half-maximum ($\Delta E$), absorption coefficient ($\alpha$) and absorption cross-section ($\sigma$) of the adjusted Gaussian curves, and their assignment to spin allowed transitions of $Cr^{3+}$ in octahedral ligand field (in *italic*) and $Cr^{4+}$ in distorded tetrahedral ligand field, for samples Cr(Ge), Cr(Ge-Al) and Cr(Al).

4 Ligand field ($Dq$), Racah parameter ($B$) and normalized ligand field ($Dq/B$) of $Cr^{3+}$ and $Cr^{4+}$ in samples Cr(Ge), Cr(Ge-Al) and Cr(Al).

**Figures :**

1 EPR spectra of the preform samples (a) Cr(Ge), (b) Cr(Ge-Al) and (c) Cr(Al)

2 Absorption spectra of the preform samples (a) Cr(Ge) and (b) Cr(Ge-Al), and (c) of the fibre sample Cr(Al). Note the strong background losses for the highly doped Cr(Ge) and Cr(Ge-Al) preforms, and its abscence for the lower doped Cr(Al) fibre.

3 Background-corrected absorption spectrum of the preform Cr(Ge) (circles), two adjusted Gaussian curves (thin solid lines) and adjusted resulting absorption spectrum (thick solid line).

4 Background-corrected absorption spectrum of the preform Cr(Ge-Al) (circles), two adjusted Gaussian curves for $Cr^{3+}$ (dashed lines), four adjusted Gaussian curves for $Cr^{4+}$ (thin solid lines), and adjusted resulting absorption spectrum (thick solid line).

5 Background-corrected absorption spectrum of the fibre Cr(Al) (circles), four adjusted Gaussian curves for $Cr^{4+}$ (thin solid lines) and adjusted resulting absorption spectrum (thick solid line).

| | Ge (mol%) | Al (mol%) | P (mol%) | Cr (mol ppm) |
|---|---|---|---|---|
| Cr(Ge) | 1.90 | 0 | 0.60 | 1400 |
| Cr(Ge-Al) | 1.15 | 0.50 | 0.50 | 600 |
| Cr(Al) | 0 | 1.00 | 0.50 | 4000 |

**Table 1**

| | EPR band position and intensity (a.u.) | | | $[Cr^{3+}]$ (mol ppm) | $[Cr^{4+}]$ (mol ppm) | $[Cr^{4+}]/[Cr_{total}]$ (%) |
|---|---|---|---|---|---|---|
| band | A | B | C | | | |
| magn. field (G) | 3420 ± 10 | 3340 ± 10 | ~1500 | | | |
| Cr(Ge) | 215 | 27 | 14 | 900 | 500 | 36 |
| Cr(Ge-Al) | 55 | 20 | 10 | 230 | 370 | 62 |
| Cr(Al) | -- | 175 | 30 | 0 | 4000 | 100 |

**Table 2**

| | $E$ ($cm^{-1}$) | $\Delta E$ ($cm^{-1}$) | $\alpha$ (dB/cm) | $\sigma$ ($10^{-24}$ $m^2$) | oxidation state | transition |
|---|---|---|---|---|---|---|
| Cr(Ge) | *14890* | *2950* | *35* | *43* | $Cr^{3+}$ | ${}^4A_2({}^4F) \rightarrow {}^4T_2({}^4F)$ |
| | *21325* | *3745* | *20* | *24* | | $\rightarrow {}^4T_1({}^4F)$ |
| Cr(Ge-Al) | 9000 | 2325 | 1.35 | 4.0 | $Cr^{4+}$ | ${}^3A_2({}^3F) \rightarrow {}^3T_2({}^3F)$ |
| | 11310 | 2340 | 1.2 | 3.5 | | $\rightarrow {}^3T_1({}^3F)$ |
| | 12310 | 2210 | 1.45 | 4.3 | | $\rightarrow {}^3T_1({}^3F)$ |
| | 16450 | 2220 | 1.85 | 5.5 | | $\rightarrow {}^3T_1({}^3F)$ |
| | *14440* | *3900* | *6.0* | *28.6* | $Cr^{3+}$ | ${}^4A_2({}^4F) \rightarrow {}^4T_2({}^4F)$ |
| | *21050* | *1630* | *4.5* | *21.5* | | $\rightarrow {}^4T_1({}^4F)$ |
| Cr(Al) | 8900 | 2000 | 0.06 | 1.6 | $Cr^{4+}$ | ${}^3A_2({}^3F) \rightarrow {}^3T_2({}^3F)$ |
| | 10970 | 3330 | 0.13 | 3.6 | | $\rightarrow {}^3T_1({}^3F)$ |
| | 13700 | 3325 | 0.135 | 3.7 | | $\rightarrow {}^3T_1({}^3F)$ |
| | 16140 | 4500 | 0.135 | 3.7 | | $\rightarrow {}^3T_1({}^3F)$ |

**Table 3**

| | $Cr^{3+}$ | | | $Cr^{4+}$ | | |
|---|---|---|---|---|---|---|
| | *Dq ($cm^{-1}$)* | *B ($cm^{-1}$)* | *Dq/B* | *Dq ($cm^{-1}$)* | *B ($cm^{-1}$)* | *Dq/B* |
| Cr(Ge) | 1490 | 660 | 2.25 | - | - | - |
| Cr(Ge-Al) | 1440 | 700 | 2.07 | 900 | 540 | 1.66 |
| Cr(Al) | - | - | - | 890 | 620 | 1.45 |
| Uncertainty | ± 10 | ± 60 | ± 0.17 | ± 10 | ± 120 | ± 0.20 |

**Table 4**

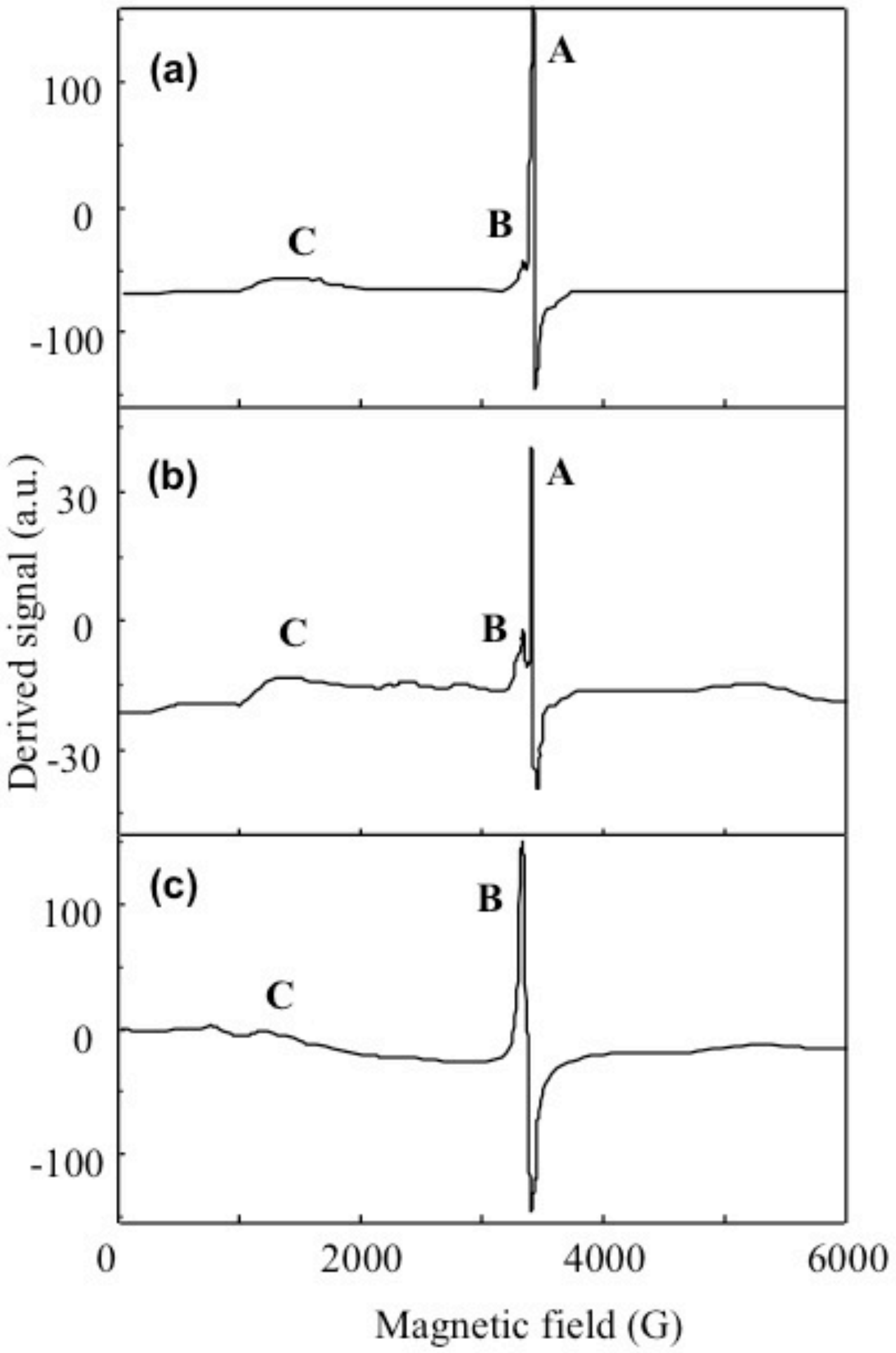

Figure 1

Felice *et al.*

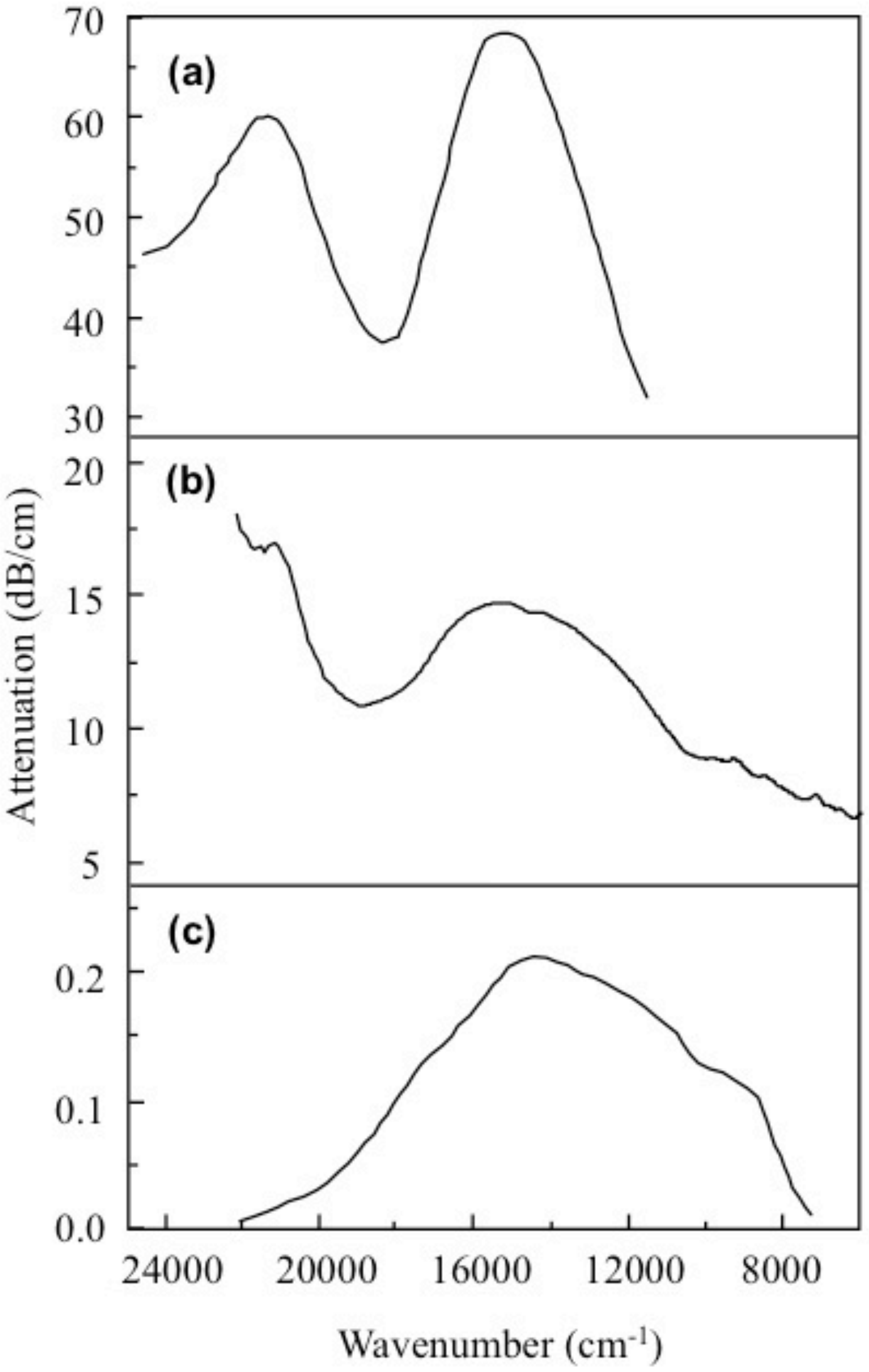

Figure 2

Felice *et al.*

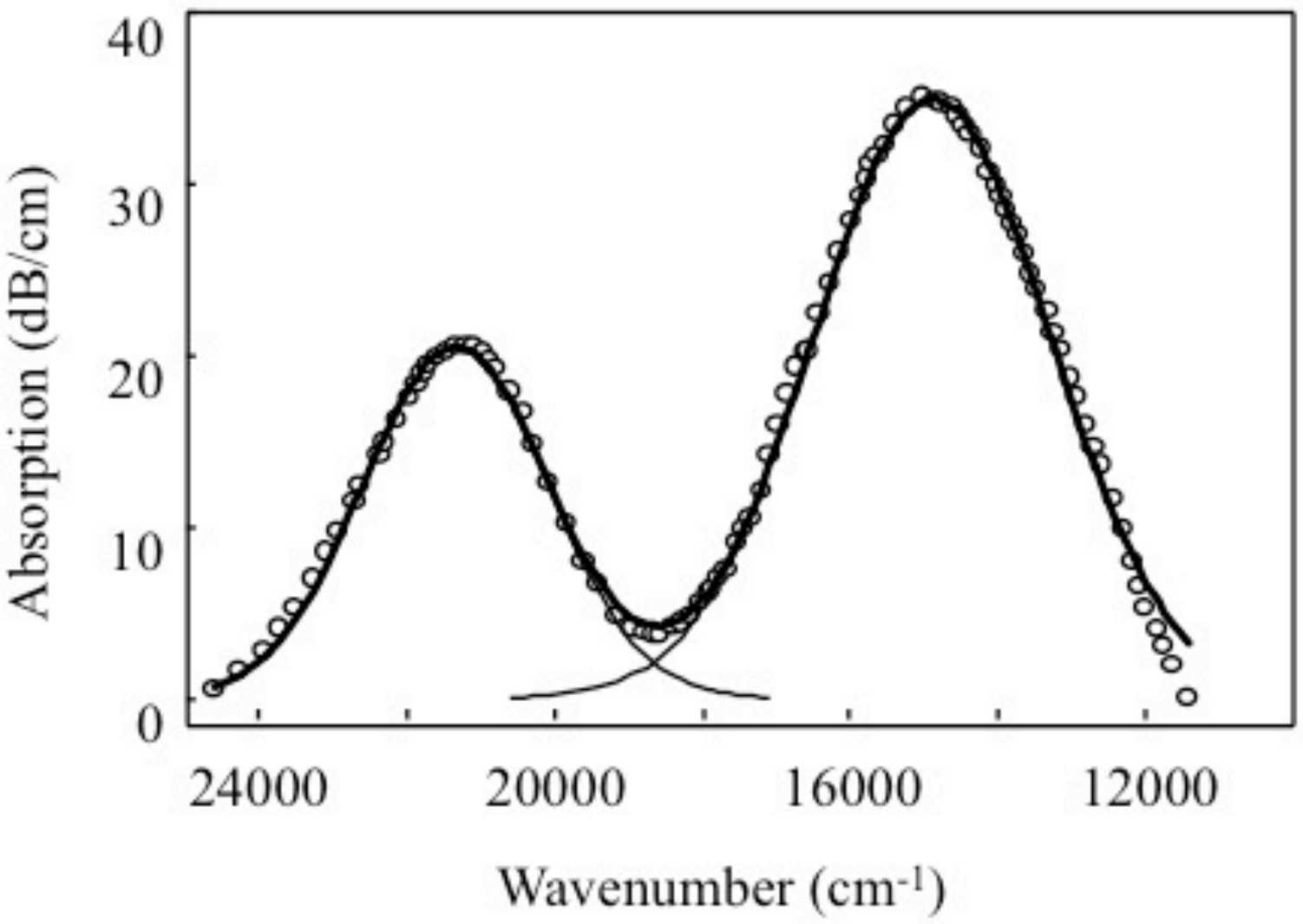

Figure 3

Felice *et al.*

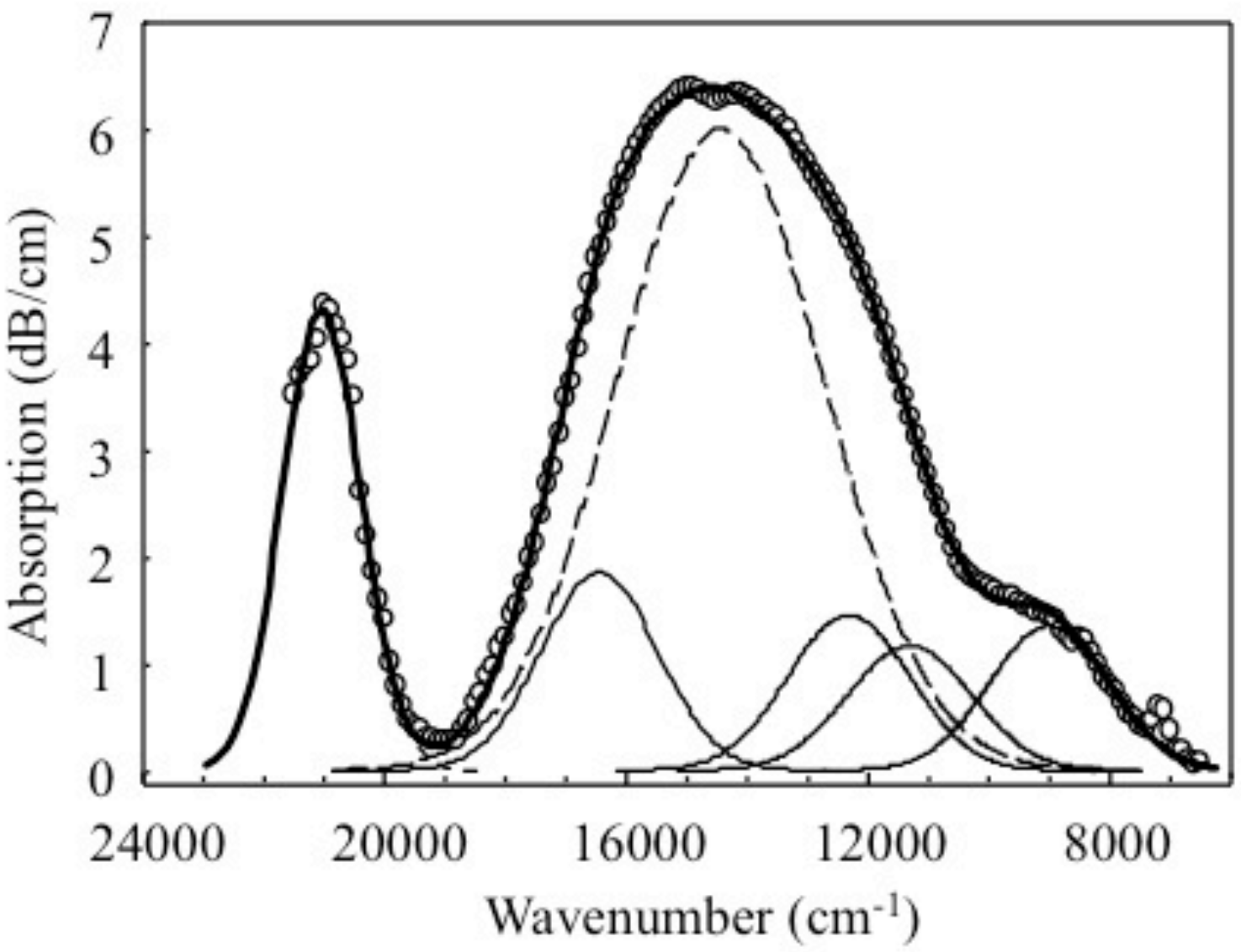

Figure 4

Felice *et al.*

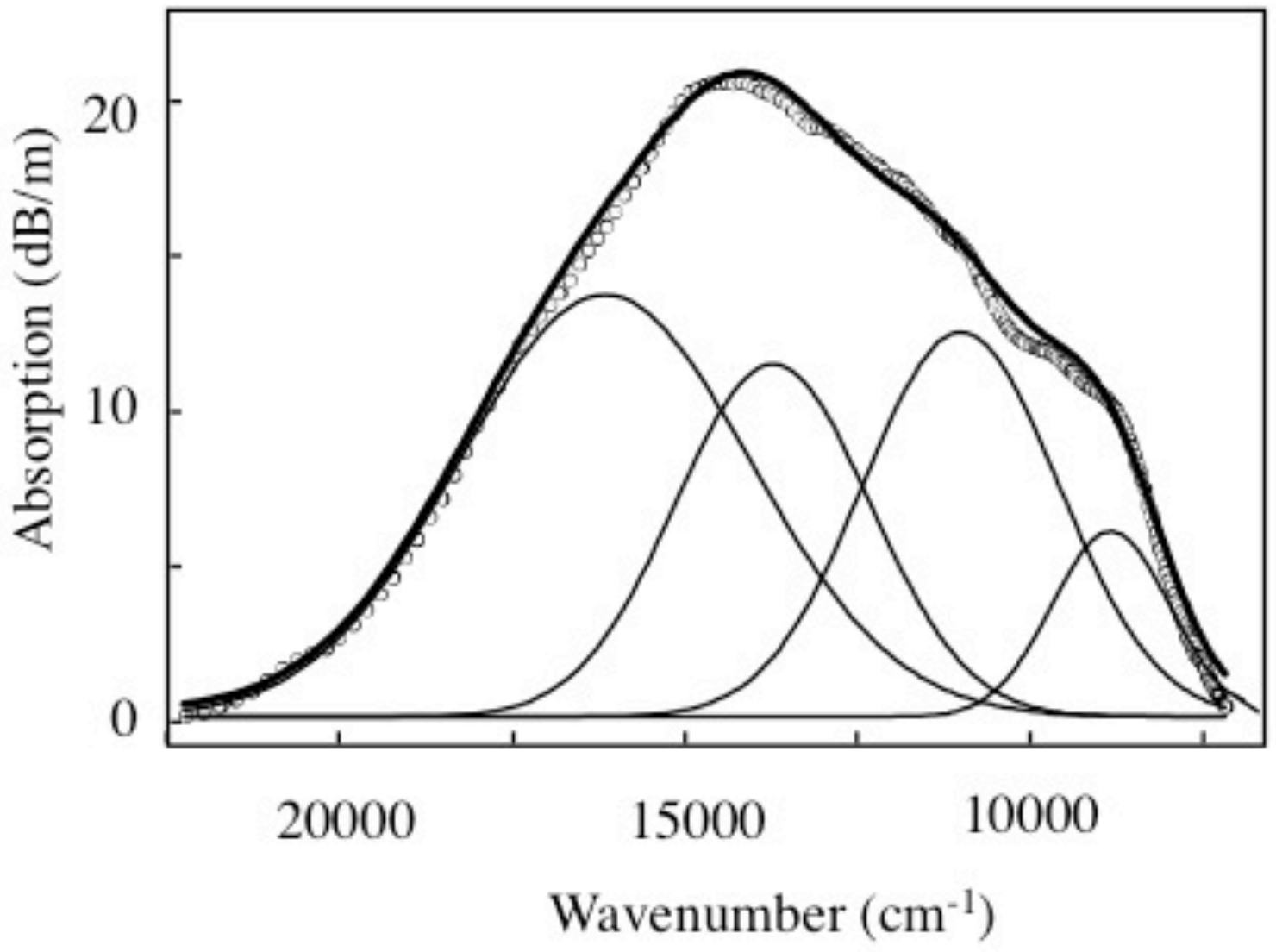

Figure 5

Felice *et al.*